\documentclass[twocolumn,prb,groupedaddress]{revtex4-1}

\usepackage[utf8]{inputenc}
\usepackage[english]{babel}
\usepackage[T1]{fontenc}
\usepackage{lmodern}
\usepackage{graphicx}
\usepackage{amsmath}
\usepackage{amsfonts}
\usepackage{amssymb}
\usepackage{microtype}
\usepackage{braket}
\usepackage{caption}
\usepackage{subcaption}

\usepackage[breaklinks=true,colorlinks, citecolor=blue]{hyperref}

\newcommand{\bi}{\bibitem}
\newcommand{\cc}{\captionsetup{justification=raggedright,singlelinecheck=false}}

\newcommand{\ct}{\cite}

\newcommand{\Tr}{\operatorname{Tr}}

\newcommand{\da}{\dagger}
\newcommand{\nn}{\nonumber}
\newcommand{\ttil}{\tilde{t}}
\newcommand{\Htil}{\widetilde{\mathcal{H}}}
\newcommand{\ktil}{\widetilde{\kappa}}
\newcommand{\Mtil}{\widetilde{M}}
\newcommand{\Ytil}{\widetilde{Y}}

\begin{document}

\title{Dynamical generation of Majorana edge-correlations in a ramped Kitaev chain coupled to  non-thermal dissipative channels }

\author{Souvik Bandyopadhyay}
\author{Sourav Bhattacharjee}
\author{Amit Dutta}
\email{dutta@iitk.ac.in}
\affiliation{Department of Physics, Indian Institute of Technology Kanpur,\\ Kanpur-208016, India}

\begin{abstract}
 We quantitatively study the out-of-equilibrium edge Majorana correlation in a linearly ramped  1D Kitaev chain of finite length in a dissipative environment. The chemical potential is dynamically ramped to drive the chain from its topologically trivial to non-trivial phase in the presence of   couplings to non-thermal Markovian baths. We consider two distinctive  situations:   In the first situation, the bath is quasi local in the site basis (local in quasi-particle basis) while in the other it is local. Following a Lindbladian approach, we compute the early time dynamics as well as the asymptotic behavior of the edge-Majorana correlation to probe the interplay between two competing time scales, one due to the coherent ramping while the other is due to the dissipative coupling. For the  quasi-local bath, we establish that there is a steady generation of Majorana correlations in asymptotic time and  the presence of an optimal ramping time which facilitates a quicker approach to the topological steady state. In the second scenario, we analyse the action of a local particle-loss type of bath in which we have established the existence of an optimal ramping time which results from the competing dynamics between the unitary ramp and the dissipative coupling. While the defect generated by the former decays exponentially with increasing ramp duration, the later scales linearly with the same.
This linear scaling is further established through a perturbation theory formulated using the non-dimensionalised coupling to the bath as a small parameter.

 \end{abstract}

\maketitle

\section{\label{sec_intro}Introduction}
 {Topological properties of out-of equilibrium quantum matter is an emerging field of both theroretical \ct{kitaev01,kane05,bernevig06,fu08,zhang08,sato09,sau10a,sau10b,lutchyn10,oreg10,oka09,kitagawa11,bermudez10,rudner13,balseiro14,mitra15,gil16,lindner11,cayssol13} and 
 experimental \ct{mourik12,rokhinson12,deng12,das12,churchill13,finck13,alicea12,leijnse12,beenakker13,stanescu13} investigations. For review, we refer to the articles  [\onlinecite{moore10,shen12,bernevig13}]). Topological properties of matter are established to be extremely robust against sufficiently weak and local time-independent perturbations \ct{kitaev01,kitaev16,pachos17}.  These robustness arises in the topological phase (e.g., of a topological insulator) that is characterised by  a non-trivial value of the topological invariant and is separated from the topologically trivial phase by a gapless quantum critical point (QCP). Further, in the topological phase of a chain with an open boundary condition, there exist a bulk-boundary correspondence (BBC) reflected  in the existence of robust zero energy edge states in its topologically non-trivial phase. More recently, the fate of such topological phases of matter and the  BBC in the presence of a {\it dissipative} environment has garnered considerable attention.}\\

 {Very promising among such topological systems is the one-dimensional (1D) p-wave topological superconductor, described theoretically by the 1D Kitaev model \ct{kitaev01,alicea12,kitaev16}. It has been established that the 1D Kitaev chain in its topological phase, hosts zero energy Majorana fermionic modes which are topologically protected. Majorana qubits being inherently non-local in character, are robust against local perturbations and are hence expected to support fault-tolerant quantum information processing operations. Non-local Majorana fermions and their statistics are theorized to find tremendous application in the implementation of fundamental quantum gates.  The experimental implementation of quantum gates in such a system however requires unitary operations to be performed on the topological Majorana fermions in the presence of environmental couplings. In this direction, there have been an upsurge in both theoretical \ct{ng15,nava19,caspel19,chun17} and experimental \ct{stevan14, gul18,ling17} studies which probe the robustness, dynamical engineering and manipulation of these topological Majorana modes under a unitary drive in the presence of dissipative environmental couplings. 
In the light of recent experimental realisation of Majorana modes in quantum nano-wires \ct{zhang18}, the robustness of these modes against coupling to a dissipative environment is fundamental to experimental quantum information processing.{We note in passing that recently a long-range version of the Kitaev chain has been proposed \ct{vodola14,huang14,vodola16,viyuela16,lepori17,dutta17} and its
topological and dynamical properties have been explored (for a review,
see [\onlinecite{maity19}]).\\

The possibility of unitarily transporting and braiding of Majorana modes in the {\it closed} Kitaev chain has been extensively explored with results indicating that  the   dynamical transport of the Majorana edge-modes through unitary annealing accross a QCP is not feasible\ct{bermudez10}  (See also, [\onlinecite{bermudez09,patel13,rajak14,dutta15}]).  This is the consequence of passage through  the gapless QCP where the zero-energy  Majorana end states mix with the bulk bands and therefore get completely delocalised into the chain without recovery. The dynamical fate of topology has also been studied in the context of an 1D Kitaev chain
in a specially engineered environment \ct{dhiel11};   {it has been observed   that a  dissipation free subspace is
dynamically generated  which can indeed preserve the equilibrium topological Majoranas  Kitaev chain in the asymptotic steady state}. {However,  the early time  dynamics of the topological Majoranas in dissipative systems and  the possibility of  protecting  unitarily prepared Majorana modes against dissipation is a largely unexplored area where our work focusses on.} \\

In this work, we consider a linearly quenched BDI symmetric  \ct{altland97} 1D finite Kitaev chain coupled to either of the  two different non-thermal baths to address the following questions: (i) How does the coupling with a quasi-local/local bath in the presence of a linear time dependent drive, affect the out of equilibrium behavior of the topological edge-Majorana correlations? (ii) Is it at all possible to engineer topological Majorana correlations in the presence of dissipation? (iii) How does coupling to the environment quantitatively affect the adiabatic preparation of correlated Majorana modes?\\

To address the issues raised in the previous paragraph, firstly we consider a {finite} 1D Kitaev chain, with an open boundary condition, coupled to a Markovian quasi-local non-thermal bath\ct{dhiel11} in the Lindbladian approach. In the presence of such a dissipator, the chemical potential is linearly ramped in time starting from a topologically trivial phase (of the closed system) to a non-trivial phase of the bare Hamiltonian.  At the end of the quench, the time dependent driving is switched off {and the system evolves with the time-independent final Hamiltonian in the same dissipative environment.} {It is noteworthy that} the dissipator is chosen such that  the Lindbladian steady state is the pure ground state of the topological Kitaev chain \ct{dhiel11}. We show numerically that although the system asymptotically reaches a steady topological state,  a short but finite ramp duration facilitates a quicker generation of edge-Majorana correlations. \\
 

In the second situation, we consider a dissipator which acts locally and independently on each site of the chain \ct{carmele15,keck17}. 
Following a similar ramp protocol as before, we exhibit the presence of an optimal ramp time for which the defect generated in the edge Majorana correlations is minimum.
The optimality results from two sources of defect generation - the non-adiabatic effects arising from finite ramp duration and the dissipative effects due to the coupling to the bath. The finite ramp duration induces defects which are suppressed exponentially with increasing ramp duration.  On the other hand, the defects induced by dissipation scale linearly with the same. 
This linear scaling has been further established using a perturbation scheme which, to the best of our knowledge, has not been reported elsewhere.\\

The paper is organised as follows: In Sec.~\ref{sec_model} we provide a brief introduction to the 1D Kitaev chain and its equilibrium topological properties. We also discuss the bulk boundary correspondence and define the defect in  the edge-Majorana correlation function along with its out of equilibrium behavior under an unitary quench across a QCP.  Further in Sec.~\ref{sec_dissi}, 
we set up the computational scheme used to analyze the edge-Majorana correlations in the presence of dissipation.
In Sec.~\ref{sec_diss_aid_prep}, we proceed with a linearly quenched Kitaev chain coupled to a quasi-local bath and probe the
the possibility of generation of correlated Majorana modes with the aid of dissipation.  In Sec.~\ref{sec_opt} we consider a local bath, which unlike the previous case is inherently detrimental to the genreration of edge Majoranas. We study the comparative time scales associated with the coherent and the dissipative dynamics and thereby identify an optimal time for which the defect generated in the edge-Majorana correlation can be minimized.
The linear scaling of the defects induced solely by dissipative coupling is validated by a perturbation theory in Sec. \ref{sec_perturb}.
We conclude in Sec.~\ref{conclude} with a brief summary of the work where  we discuss relevant connections with recent experiments and the scope of further research.  {Three}
Appendices have been incorporated to complement the discussions presented in the main text.

\section{\label{sec_model}Kitaev chain and unitary dynamics across a QCP}
\subsection{Topological properties}
The Kitaev chain is a one-dimensional system of spinless fermions on a lattice of linear dimension $L=Na$, where $N$ is the number of sites and $a$ is the lattice spacing which we henceforth set equal to unity. The
model is represented by the {many-body} Hamiltonian \ct{kitaev01}
\begin{align}\label{eq_hamil_r}
H=-\sum_{n=1}^{N-1}\left(Jc_n^\dag c_{n+1} +\Delta c_nc_{n+1} + h.c.\right)\nn\\
-\mu\sum_{n=1}^N\left(2c_n^\da c_n-1\right);
\end{align}
note that we have set $\hbar=1$ throughout and shall use the natural unit system in which length and time are on the same footing. The first term on the r.h.s of Eq.~\eqref{eq_master} captures the unitary part of the evolution driven by the time dependent Hamiltonian $H(t)$. The second term, which comprises of the coupling constants $\kappa_j>0$, and Lindblad operators $L_j$, represents the non-unitary dissipative part of the evolution. The choice of the Lindblad operators considered in this work will be discussed in later sections.
 addition to the chemical potential $\mu$ and the nearest-neighbor (NN) hopping interaction of amplitude $J$, the Hamiltonian also incorporates an additional NN pairing interaction of amplitude $\Delta$. Naturally, the total number of fermions is not conserved, even though parity is. The bulk of the chain located away from the edges can be modelled as a closed chain with periodic boundary conditions (PBC). Within the bulk, translation invariance allows a Fourier transformation of the bulk Hamiltonian into the (quasi) momentum basis where it assumes a particularly convenient form, 
\begin{align}\label{eq_hamil_k}
H=\bigoplus_{k>0}H_k=\bigoplus_{k>0}\left(H_k^e\bigoplus H_k^o\right)-(2J\cos{k})\mathcal{I}_k,
\end{align}
where $H_k^e=\vec{h}_k\cdot\vec{\sigma}_k$ and $H_k^o=\mathbb{O}_{2\times 2}$ are the decoupled {single-particle} Hamiltonians of even and odd parity sectors of each momentum mode {derived
from the many body Hamiltonian \eqref{eq_hamil_r}}, respectively. $H_k^e$ is defined in  terms of  Pauli matrices $\vec{\sigma}_k$ and a vector $\vec{h}_k$ with components
\begin{subequations}\label{eq_hamil_k_comp}
\begin{align}
&h_k(x)=0\label{eq_hx}\\
&h_k(y)=-2\Delta\sin{k}\label{eq_hy}\\
&h_k(z)=2\mu+2J\cos{k}\label{eq_hz}.
\end{align}	
\end{subequations}
The energy eigen values of $H_k^e$ are given by 
\begin{align}\label{eq_spec}
E_k=\pm\sqrt{h_k(y)^2+h_k(z)^2}.
\end{align}
The ground state of $H_k$ (and consequently $H$) therefore lies in the even parity sector. We shall assume $\Delta$ to be real so that $H_k$ is completely BDI symmetric. Therefore, the  topological phase
is characterised by an integer-quantised winding number \ct{kitaev16} for  periodic boundary conditions and separated from the trivial phase (with zero winding number) through gapless quantum critical points.
The non-zero winding number physically manifests itself in the form of a BBC i.e.,  the ground state of the open chain supports localised  zero-energy 
Majorana end modes in the thermodynamic limit. These zero energy modes, residing in the mid-gap of the bulk spectrum, are consequently robust against weak local perturbations.

\begin{figure}
	\includegraphics[width=\columnwidth]{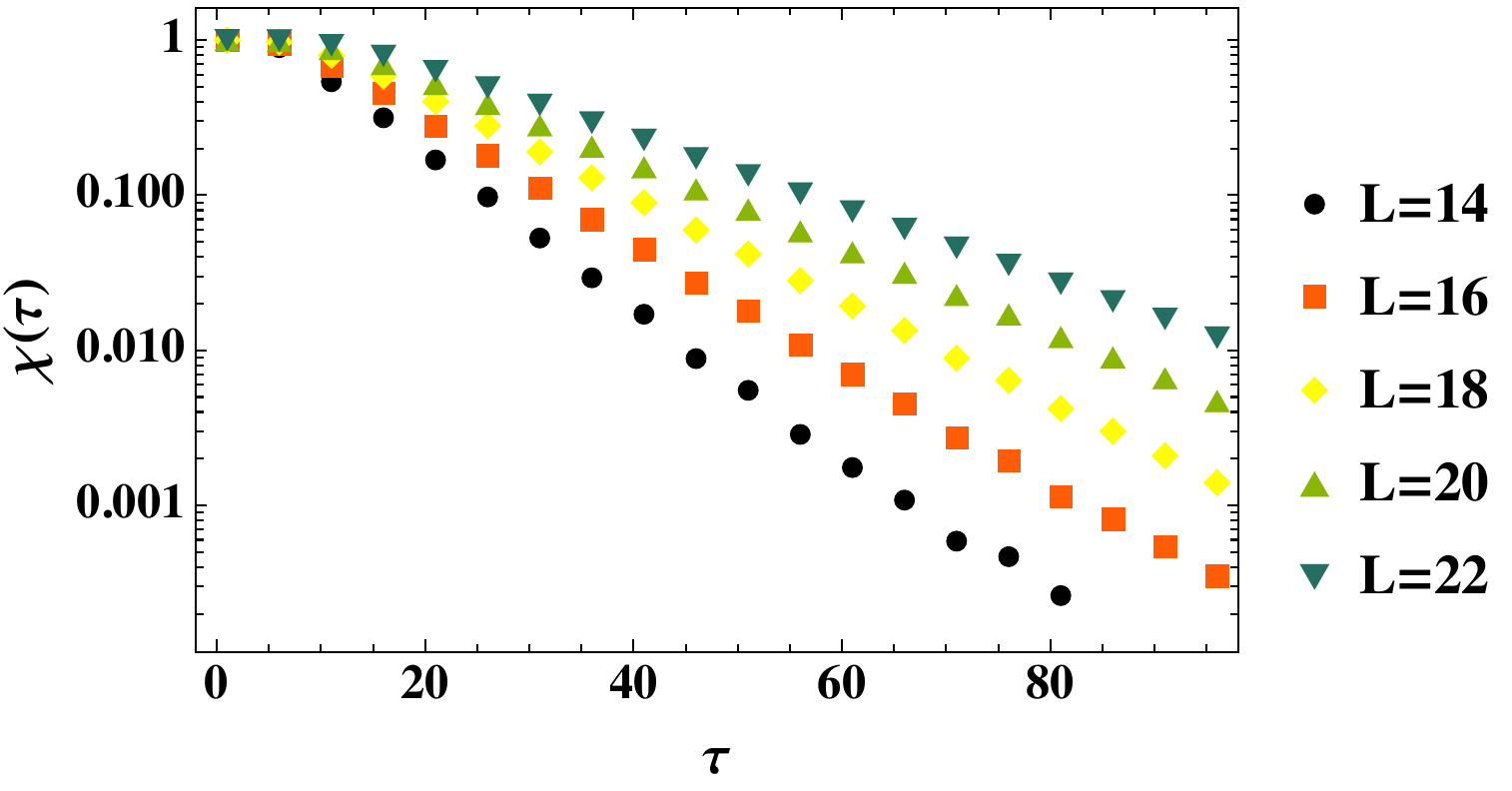}	
	\cc
	\caption{Defects in two-point correlation of Majorana edge modes (see Eq.~\eqref{eq_defect}), {measured at time $t=\tau$}, generated in the case of purely unitary ramping (defined in Eq.~\eqref{eq_protocol}) as a function of ramp duration $\tau$ for varying system  size $L$. The linear scaling of the defects when plotted in semi-log scale suggests that they scale as $\chi(\tau)=e^{-f(L,\mu_i)\tau}$ with $f(L,\mu_i)\geq 0$. The relevant parameters chosen are $J=1$, $\mu_i=2$ and $\mu_f=0$.}
	\label{fig_unitary}
\end{figure} 
 
To further elucidate the BBC, the Hamiltonian in Eq.~\eqref{eq_hamil_r} is mapped to the Majorana system through the transformation $c_n=(a_{2n-1}-ia_{2n})/2$ where $a_n$ are the self-adjoint Majorana operators. Here and henceforth, we set $\Delta=-J$ for purpose of simplicity and assume that both $\mu$ and $J$ are real. The transformed Hamiltonian assumes the form
\begin{align}\label{eq_hamil_ma}
H=i\mu \sum_{n=1}^{N}a_{2n-1}a_{2n}-iJ\sum_{n=1}^{N-1}a_{2n}a_{2n+1}.
\end{align}
Further, an appropriate Bogoliubov transformation diagonalizes the Hamiltonian in Eq.~\eqref{eq_hamil_r} as
\begin{align}\label{eq_hamil_diag}
H=\sum_{i=1}^NE_i(d_i^\da d_i-d_i d_i^\da)=E_g+2\sum_{i=1}^NE_id_i^\da d_i,
\end{align}
where $E_g=-\sum_iE_i$ is the ground state energy of the system and corresponds to the \textit{Bogoliubov vaccum} $\ket{GS}$ ($d_i\ket{GS}=0$, $\forall i$) in which all the negative energy states are occupied, while $E_i$ represents the energy of quasi-particle excitations generated by the Bogoliubov fermionic creation operators $d_i^\da$ acting on the the ground state. In other words, the operators $d_i$ annihilates   the Bogoliubov vacuum. The BBC is now explicitly identified as follows: for $|\mu|<|J|$, the bulk winding  number is quantized to unity; correspondingly, the two-point correlation function of the Majorana end modes in the ground state of the Hamiltonian, defined  as {$\bra{GS}\theta\ket{GS}$, where $\theta=i a_1a_{2N}$}, also remains finite and approaches unity as $\mu\to 0$. The localization of the  Majorana modes stem from the presence of a zero-energy quasi particle excitation ($E_z=0$, $E_z\in\{E_i\}$) in the thermodynamic limit. On the other hand, for $|\mu|>|J|$, $\theta$ vanishes in the thermodynamic limit in the trivial phase. The two phases are demarcated by a quantum critical point (QCP) at $|\mu|=|J|$; at this point, the bulk spectrum becomes gapless in the thermodynamic limit.

\subsection{Unitary dynamics across a quantum critical point}\label{subsec_unitary}
The presence of a QCP or vanishing gap in the bulk spectrum presents a conundrum in the context of  preparing a topological state --- starting from a trivial phase of the system, it is impossible to drive the system into a non-trivial phase through a unitary dynamics \ct{bermudez10}. Specifically, in the problem that we consider in this work, the system is initially in the ground state $\ket{\psi(0)}=\ket{\psi_i^0}$ of an initial Hamiltonian $H_i: |\mu_i|>|J|$, following which the Hamiltonian is ramped across the QCP at $|\mu|=|J|$ to a final $H_f:|\mu_f|<|J|$ using the protocol,
\begin{align}\label{eq_protocol}
\mu(t)=\left(\mu_i+(\mu_f-\mu_i)\frac{t}{\tau}\right)\Theta(\tau-t)+\mu_f\Theta(t-\tau),
\end{align}
where $\Theta(x)$ is the Heaviside step function. The protocol therefore linearly ramps the initial $\mu_i$ to a final $\mu_f$ during time $\tau$ after which $\mu$ remains frozen at the targeted $\mu_f$. In the thermodynamic limit, the quantum adiabatic theorem breaks down; the system $\ket{\psi(t)}$ therefore cannot be exclusively prepared in the ground state $\ket{\psi_f^0}$ of $H_f$. This results in generation of \textit{defects} in the correlation of the Majorana end modes which we quantify as 
\begin{align}\label{eq_defect}
\chi(t)=\bra{\psi_f^0}\theta\ket{\psi_f^0}-\bra{\psi(t)}\theta\ket{\psi(t)}.
\end{align}
We remark that although the ground state $\ket{\psi_f^0}$ is doubly degenerate in the thermodynamic limit, we choose the state $\ket{\psi_f^0}$ as the Bogoliubov vacuum corresponding to the final Hamiltonian. The choice of the ground state however, do not effect the results qualitatively. 
Note that for the ideal situation of  a perfect unitary adiabatic preparation, the quantity $\chi(t)$ should vanish.

However, for a finite system of size $L$, the bulk spectrum is not truly gapless; the gap $\delta$ at the QCP scales as $\delta\sim 1/L$. Consequently, if the time-scale $\tau$ of the ramp is large enough such that $\tau\gg L$, the bulk of the chain (closed chain with PBC) can be prepared in the ground state of the final Hamiltonian. However, unlike the bulk, the splitting between the edge modes vanishes exponentially with $L$. Therefore, even for a finite system, the defect $\chi(t)$ truly vanishes  only in the limit $\tau\to\infty$. Nevertheless,  it is possible for $\chi(t)$ to arbitrarily approach zero for a finite size systems even for finite $\tau \gg L$. This is illustrated in Fig.~\ref{fig_unitary} where we also show that  the rate at which $\chi(\tau)$ approaches zero decrease with increasing system size. The key point of this section is that perfect Majorana correlations cannot be generated if the ramp duration is finite with  the defects scaling as $\chi(\tau)=e^{-f(L,\mu_i)\tau}$, where the coefficient $f(L,\mu_i)\geq 0$ is non-universal. 

We note in passing that a similar (dissipationless) preparation of edge Majorana correlations in an extended version of the Kitaev chain was studied in Ref.~[\onlinecite{kraus12}].


\section{Dissipative dynamics of edge modes}\label{sec_dissi}
In this section, we address  the dynamics of the Majorana edge modes when the Hamiltonian is ramped in the presence of a dissipative environment.  The purpose of this section is to set up the computational scheme that has been used to obtain the results discussed in subsequent sections. Let us assume that the dynamical evolution of the system state, represented by the density matrix $\rho(t)$, can be described through a differential equation of the following Lindblad form \ct{breuer02}: 
\begin{align}\label{eq_master}
\frac{\partial\rho}{\partial t}=-i[H(t),\rho(t)]+\sum_j\kappa_j\left(2L_j\rho(t)L_j^\da-\{L_j^\da L_j,\rho(t)\}\right);
\end{align}

\begin{figure*}
\centering
\begin{subfigure}{0.4\textwidth}
	 \centering
	   \includegraphics[width=\columnwidth]{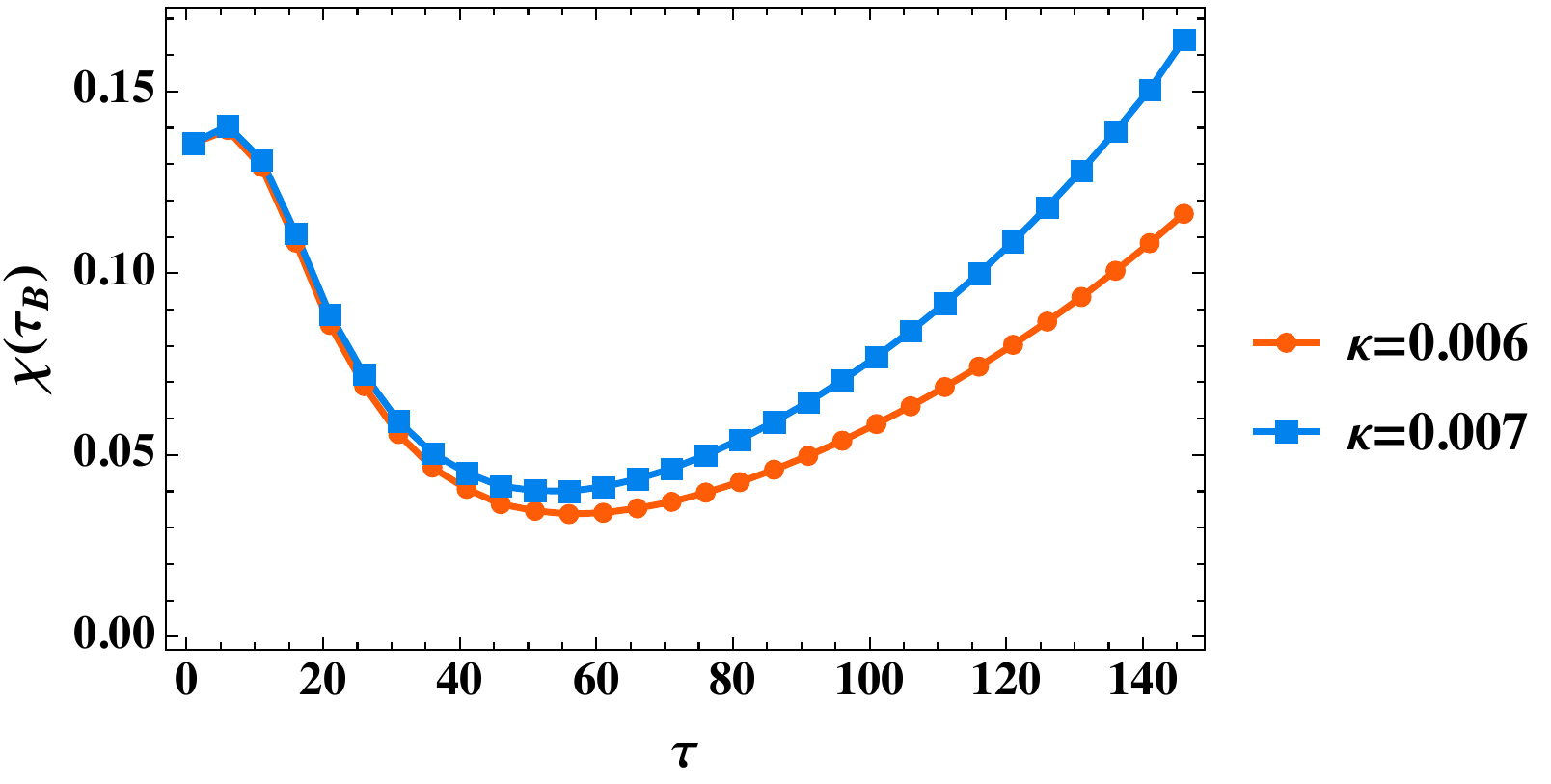}
	  \caption{}	\label{fig_2a}
	   \end{subfigure}\quad\quad\quad\quad
	   \begin{subfigure}{0.4\textwidth}
	   	\centering
	   \includegraphics[width=\columnwidth]{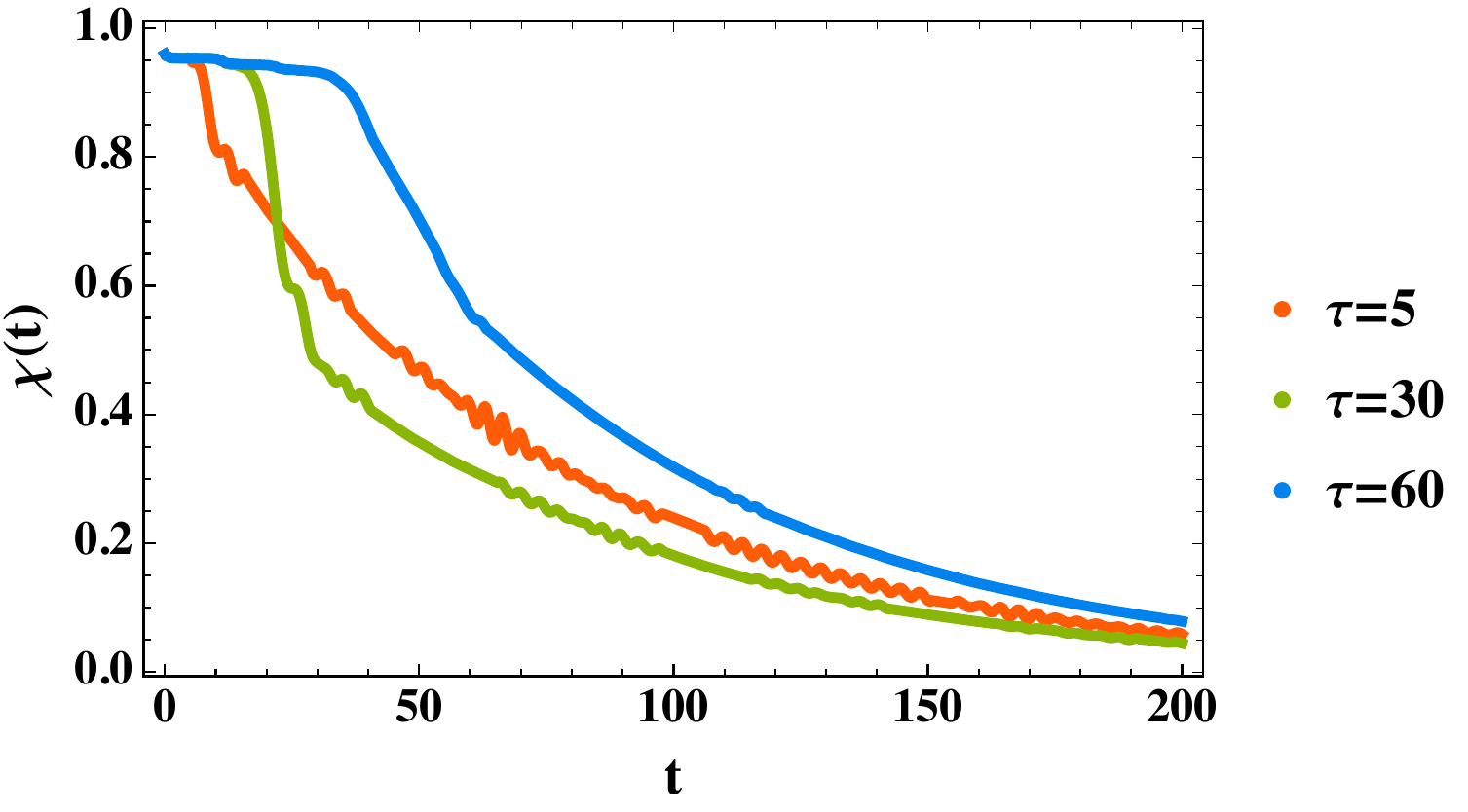}
	  	\caption{}
	   	\label{fig_2b} 
	    \end{subfigure}
				
	\cc
	\caption{ (a) Defects generated at  time $t=\tau_B$ as a function of ramp duration $\tau$  for different coupling strengths with a bath characterized by the Lindblad operators defined in Eq.~\eqref{eq_lindblad_bogo}. A small but non-zero ramp duration results in lowering of the defect $\chi(\tau_B)$. Increasing the coupling strength also lowers $\chi(\tau_B)$ signifying that the bath induces the asymptotic preparation of localized Majorana edge modes. The parameters chosen for numerical simulation are $L=20$, $J=1$, $\mu_i=2$ and $\mu_f=0$. { The lowering of defect for a finite ramp duration is also apparent in (b) where we plot the real time generation of defect $\chi(t)$ as a function of time $t$ for different ramp durations $\tau$. The  defect  generated at sufficiently large times is lowered as we move away from the sudden quench limit $\tau=5\to\tau=30$ but again increases as the ramp duration increases from $\tau=30\to\tau=60$. The parameters chosen are $L=20$, $J=1$, $\mu_i=2$, $\mu_f=0.2$, and $\kappa=0.007$.} } 
	\label{fig_bogo_tauB}

\end{figure*}

The scheme of our analysis is as follows. At time $t=0$, the system is initially prepared in the topologically trivial  ground state of an initial Hamiltonian $H_i$ such that $|\mu_i|>|J|$. The initial state $\ket{\psi_i^0}$  is subsequently allowed to evolve under a linear ramp of the chemical potential $\mu$ with the ramp protocol outlined in   Eq.~\eqref{eq_protocol} in the presence of environmental coupling. The ramp is terminated at a final $|\mu_f|>|J|$ so that the system would have been prepared in the topologically non-trivial state with finite edge Majorana correlations $\theta$ if the ramp were  dissipation-less and adiabatic  ($\kappa=0$, $\tau\to\infty$). Note that the dissipation continues to act even after the ramp is complete. Under non-unitary dynamics, the system in general is expected to be in a mixed state; the defect $\chi(t)$ in edge mode correlations is accordingly modified as (compare with Eq.~\eqref{eq_defect}) 
\begin{align}
	\label{eq_chi_mix}
\chi(t)=\bra{\psi_f^0}\theta\ket{\psi_f^0}-\Tr\big(\rho(t)\theta\big),
\end{align}
where, the evolution of the  non-equilibrium state $\rho(t)$ of the system is governed by Eq.~\eqref{eq_master}. The defect $\chi(t)$, which lies within the range $0\leq\chi(t)\leq1$, is calculated numerically (see Appendix~\ref{app_num}) and is the principle quantity of interest in the rest of our analysis.



\section{\label{sec_diss_aid_prep}Dissipation aided preparation of Majorana edge modes}

In this section, we consider the dissipative dynamics of the Majorana edge modes with the Lindblad operators in Eq.~\eqref{eq_master} chosen as
\begin{align}\label{eq_lindblad_bogo}
L_j=d_j^f,
\end{align}
where $d_j^f$ are the Bogoliubov annihilation operators defined in Eq.~\eqref{eq_hamil_diag}. Further, the operators $d_j^f$ annihilate the ground state of the final Hamiltonian $H_f$, i.e. $d_j^f\ket{\psi_f^0}=0$, $\forall j$ (hence the additional superscript $f$). We also assume that all the Lindblad operators act uniformly on the system ($\kappa_j=\kappa$, $\forall j$). Following the scheme outlined in Sec.~\ref{sec_dissi}, we proceed to analyze the defect generated in the edge mode correlations $\chi(t)$.
 
Let us first consider the asymptotic steady state of the system $\rho_{ss}=\lim_{t\to\infty}\rho(t)$ . As the system evolves under the action of the constant Hamiltonian $H_f$ for $t>\tau$, the asymptotic steady state can therefore be found by substituting $H(t)=H_f$ in Eq.~\eqref{eq_master} and equating the r.h.s to zero,  
\begin{align}\label{eq_steady}
-i[H_f,\rho_{ss}]+\kappa\sum_j\left(2d_j^f\rho_{ss}d_j^{f\da}-\{d_j^{f\da} d_j^f,\rho_{ss}\}\right)=0.
\end{align}
Solving the above equation, one obtains $\rho_{ss}=\ket{\psi_f^0}\bra{\psi_f^0}$ (see Appendix.~\ref{app_steady}). Hence the system asymptotically approaches the topological ground state of  the final Hamiltonian $H_f$. Naturally, $\lim_{t\to\infty}\chi(t)$ vanishes as can be seen by substituting $\lim_{t\to\infty}\rho_t=\rho_{ss}$ in Eq.~\eqref{eq_chi_mix}. The dissipative environment therefore induces the preparation of localized edge Majorana modes.

For a dissipative evolution governed by Eq.~\eqref{eq_master} with a time independent Hamiltonian, the time-scale of relaxation to steady state  {is of the order $\tau_B\sim1/\kappa$} {(see Appendix.~\ref{app_num})}. 
 {Altough the bath asymptotically drives the system to its pure topological state, 
 the non-equilibrium state of the system is neccessarily mixed. This is reflected in the increasing deviation of the edge-correlation function from its unitary value with increasing dissipation strength.} 


In Fig.~(\ref{fig_bogo_tauB}) we observe that a finite but small ramp duration ($0<\tau\ll\tau_B$) results in generation of lesser defects in the edge mode correlations after time $\tau_B$.  {Hence, the non-equilibrium state has a higher fidelity to the asymptotic steady state at $t=\tau_B$. This in turn is expected to  facilitate a quicker stabilization of the edge correlation into its topological steady value}. {It is
note-worthy that our result holds true even when $\mu_f \neq 0$ as elaborated in Fig.~(\ref{fig_2b}) .}

The lowering of defects (for $\tau\lesssim\tau_B$) is a consequence of the following -- (i) within the ramp duration, the bath induces negligible dissipation, the dynamics is therefore dominated by the unitary ramp. This results in lesser generation of defects at the end of the ramp with increasing ramp duration (see Fig.~(\ref{fig_unitary})). (ii) The short duration of the ramp (having $\tau\ll\tau_B$) does not significantly alter the  {dissipative relaxation time scale of the system. In other words, after the ramp is switched off, the non-equilibrium state develops a high fidelity with the topological state. This in turn is expected to result in a faster dissipative relaxation into the topological steady state. We therefore conclude with the key result that  a finite but short ramp duration  speeds up the preparation of edge modes using a bath which is local in the quasi-particle basis.
  \begin{figure*}
  	\centering
  	\begin{subfigure}{0.4\textwidth}
  		\centering
  		\includegraphics[width=\columnwidth]{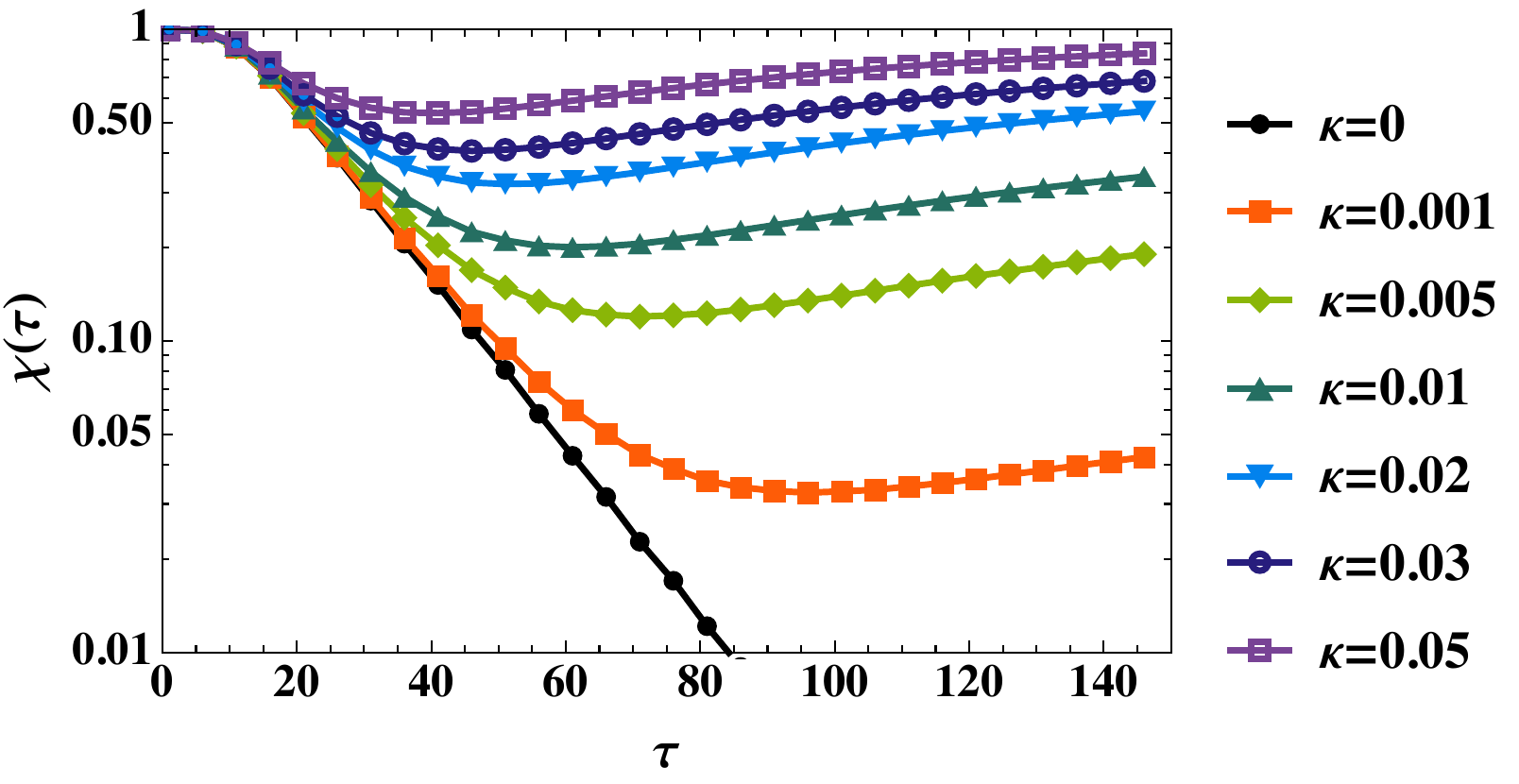}
  		\caption{}
  		\label{fig_3a}
  	\end{subfigure}\quad\quad\quad\quad
  	\begin{subfigure}{0.4\textwidth}
  		\centering
  		\includegraphics[width=\columnwidth]{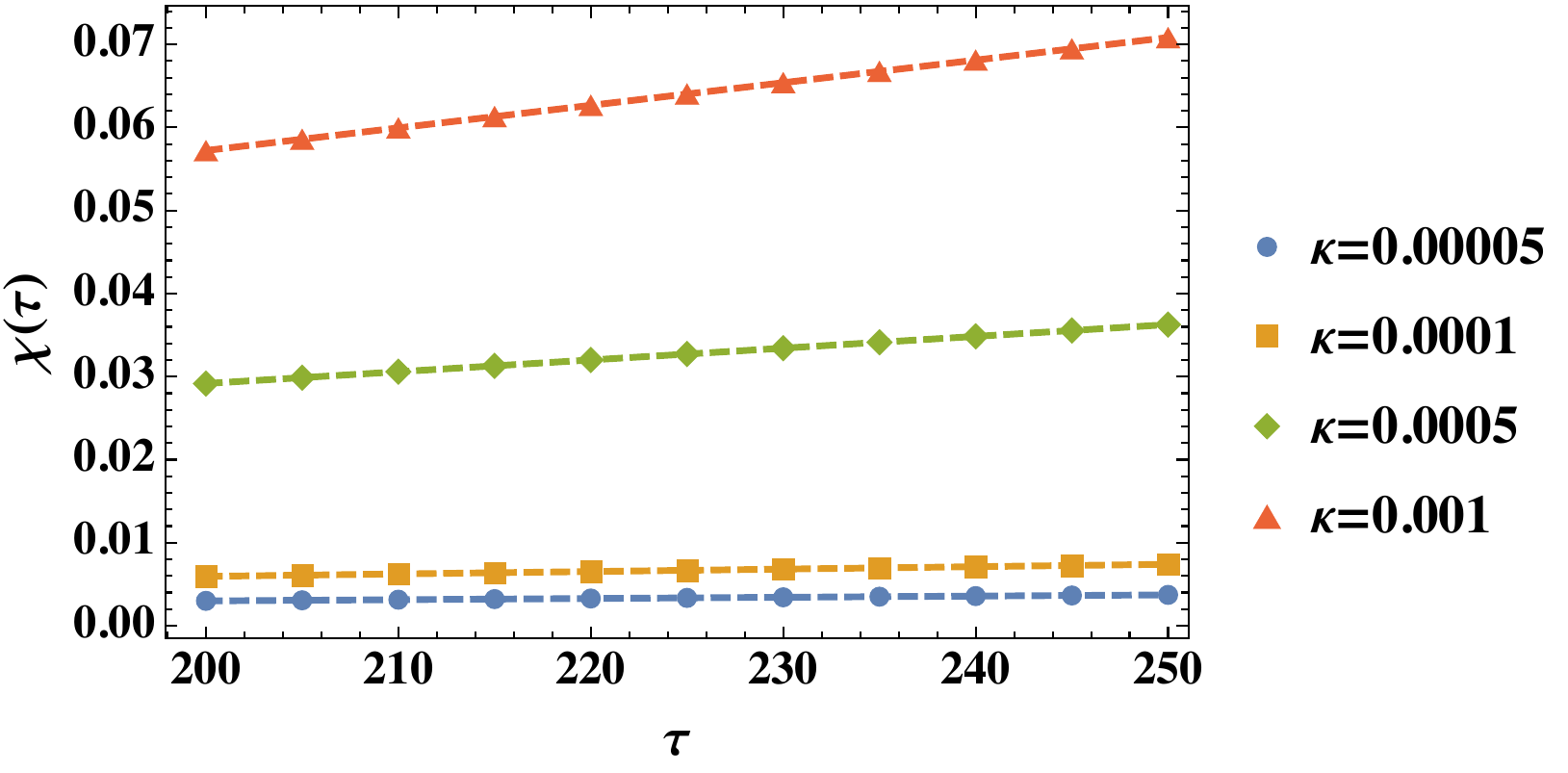}
  		\caption{}
  		\label{fig_3b} 
  	\end{subfigure}
  	\cc	
  	\caption{(a) {Defects generated at the end of the ramp $\chi(\tau)$ as a function of $\tau$ for a local bath with Lindblad operators of form Eq.~\eqref{eq_lindblad_local}. For $\tau<L\lll\tau_B$, the dynamics is dominated primarily by the unitary evolution thereby resulting in lowering of defects with increasing $\tau$. For $L\ll\tau<\tau_B$, the ramp is adiabatic and induces negligible defects on its own. The defects generated is then primarily through dissipative effects resulting in more defect generation with increasing $\tau$. The interplay of the unitary and dissipative dynamics of the bath results in an optimal $\tau_o$ for which the defect generated is minimum. (b) For $L\ll\tau\ll\tau_B$, the defect scales as $\chi(\tau)\sim\kappa\tau$. The dashed lines correspond to linear fits of the data points. In both (a) and (b), the relevant parameters chosen  are $L=20$, $J=1$, $\mu_i=2$ and $\mu_f=0$.}}
  \end{figure*}
\section{Optimality in the  presence of local dissipation}\label{sec_opt}
In this section, we model the dissipative effects through local Lindblad operators that act locally at each site on the Kitaev chain, barring the edge sites. Specifically, we choose
\begin{align}\label{eq_lindblad_local}
L_j=c_j, \quad j\in\{2,3...,N-1\}, 
\end{align} 
where $c_j$ are the fermionic annihilation operators acting on site $j$ of the chain. In the presence of such local dissipative channels, the Majorana edge modes which were initially present in the system, are known to decay exponentially in time \ct{carmele15} due to the finite overlap of the edge modes with the bulk for $\mu\neq0$. However, for $\mu=0$, the edge modes are essentially disconnected from the bulk and are therefore robust against any dissipation induced in the bulk.  In our protocol, we therefore set $\mu_f=0$, so that the edge modes once created (with some defects) remain localized after the ramp, i.e.  $\chi(t>\tau)=\chi(\tau)$. This allows us to focus on the dissipative effects on the preparation stage of localized edge modes only and exclude any defects generated after the ramp is complete. 

The key observations from the numerical results are two-fold. Firstly, unlike the previous case discussed in Sec.~\ref{sec_diss_aid_prep}, the presence of local dissipative channels in the bulk is always detrimental to the preparation of edge states, as can be seen from Fig.~\ref{fig_3a}. The second and more significant result is that there exists an optimal ramp duration for which the defects generated are minimum, which, as
we shall elaborate below, arises due to the competition between the unitary and the dissipative dynamics.

 To comprehend these results, it is instructive to compare the three relevant length/time scales in the dynamics -- $L$, $\tau$ and $\tau_B$. Assuming a weak coupling strength $\kappa$ ($1/\tau_B$)  amounts to setting $L$, $\tau$ $\ll\tau_B$. In the adiabatic limit of the ramping protocol, i.e.  $L\ll\tau\ll\tau_B$, the defects generated scales as $\chi(\tau)\sim\kappa\tau$ (see Fig.~\ref{fig_3b}). Intuitively, this monotonic rise in defect can be explained as follows --- as the edges interact with the bath indirectly through the bulk, an increase in ramp duration $\tau$ implies that the edge modes have proportionally increasing time to decay  before the chemical potential is eventually ramped to $\mu_f=0$ at $t=\tau$, following which the edge modes can no longer decay. 
In the next section, we introduce a perturbative expansion (with $\kappa\tau\ll 1$ as small parameter) of  the solution of the dynamical equations of motion for the two-point Majorana correlations to show that any observable, which is a linear function of two point Majorana-correlations, is indeed expected to  follow the same linear scaling $\sim \kappa\tau$  under the action of linear Lindblad operators.

On the other hand,  {for very fast quenches $\tau<L\lll\tau_B$}, the chemical potential $\mu$ is quickly ramped to zero within a short duration $\tau$; within this duration the environment fails to induce any substantive decay in the Majorana edge correlations. The short duration of the quench however, itself results in the generation of defects as discussed in Sec.~\ref{sec_model}. The defects scale with the ramp duration as $\chi(\tau)\sim e^{-f(L,\mu_i)}$ (see Fig.~\ref{fig_unitary}), where the coefficient $f(L,\mu_i)\geq 0$ is a model dependent non-universal function which can not be derived within any analytical framework. The defect generation in the fast quench limit is therefore dominated by the unitary dynamics and arises due to the fast (non-adiabatic) ramping.

It follows that, there exists an \textit{optimal} ramp duration $\tau_o$ at which the defect generation {at the end of the ramp} is minimized (see Fig.~\ref{fig_3a}). From the preceding discussions, one can assume that the defect generated for $\tau\sim\tau_o$, has the form as $\chi(\tau)=\exp\left(-f(L,\mu_i)\tau\right)+\kappa\tau$. A generic expression for $\tau_o$ can be derived by minimizing the the defect with respect to $\tau$ as,
\begin{align}
\frac{d}{d\tau}\left(\exp\left(-f(L,\mu_i)\tau\right)+\kappa\tau\right)|_{\tau=\tau_0}=0
\end{align}
or,
\begin{equation}
\tau_o=-\frac{1}{f(L,\mu_i)}\log\left(\frac{\kappa}{f(L,\mu_i)}\right).
\end{equation}
The positivity of $f(L,\mu_i)$ along with the condition of weak-coupling strength $\kappa$ implies that there exist a positive definite $\tau_o$ at which the defects are minimised. The existence of the optimal $\tau_o$ is a consequence of the competition between the unitary dynamics which demands a large $\tau$ for defect minimisation and the dissipative effects which requires short $\tau$ for the same.

We emphasise here that the above expression for the optimal ramp duration is not universal because the scaling of the defects arising from non-adiabatic effects is model dependent as reflected in the coefficient $f(L,\mu_i)$. This is unlike the universal scaling of the optimal ramp time obtained in the case of defect generation in residual energy \cite{keck17}, where the residual energy is defined as the excess energy of the time-dependent state over the instantaneous ground state and is quantitatively obtained by replacing $\theta$ in Eq.~\eqref{eq_chi_mix} with $H(t)$ so that  $H(\tau)=H_f$. {Consequently, the residual energy is an extensive (bulk) property of the system}. The universality in the scaling of residual energy stems from the fact that the contribution to the defects from the non-adiabatic excitations follow a universal Kibble-Zurek scaling.

 {
We conclude this section with the remark that Ref. [\onlinecite{carmele15}] established that edge Majorana correlations initially present in the system do not survive when the local dissipative coupling is turned on. On the
other hand, our results show that an identical dissipative environment is also detrimental to the preparation of the same Majorana edge correlation through a linear ramping protocol starting from a trivial phase. Additionally, it was also argued in Ref. [\onlinecite{carmele15}] that in the $\mu=0$ case, perfect Majorana correlation is preserved as the edge Majorana modes are completely decoupled from the bulk and hence from the bath. Our results however show that preparation of such perfect Majorana correlation is impossible even with $\mu=0$ in the final Hamiltonian, as the Majorana modes interact with the bath indirectly through the bulk during the finite duration of the ramp.  No optimality in the behaviour of the Majorana correlation was obtained in Ref [56] as the only relevant scale in the system arose from the relaxation dynamics due to the dissipative bath. The presence of a linear ramp in our work incorporates an additional time-scale; the optimality observed  is an artefact of these two competing time-scales.}

{\section{Perturbative Approach}\label{sec_perturb}
While the lowering of defects with increasing ramp duration $\tau$ (in the small $\tau$ limit) is a result of approach towards the adiabatic limit of ramping protocol, the linear rise following the optimal $\tau_0$ needs further scrutiny. 

To this end, we pertubatively expand the solution of the dynamical equations of motion for the two-point Majorana correlations where we make use of the condition that $\tau\ll\tau_B$ or $\kappa\tau\ll 1$ ($\kappa\sim1/\tau_B$) to identify $\kappa\tau$ as the small parameter in the perturbation.}

{The dynamical equations of motion for the two-point Majorana correlations can be expressed in terms of a $2L\times2L$ dimensional covariance matrix $C(t)$ defined as $C_{i,j}(t)=\Tr(a_ia_j\rho(t))-\delta_{i,j}$, which satisfies (see Appendix.~\ref{app_num})
\begin{align}
\dot{C}(t)=-X(t)C(t)-C(t)X^T(t)+iY(t),
\end{align}
where, $X(t)=4(i\mathcal{H}(t)+\mathrm{Re}[M(t)])$, $Y(t)=4(\mathrm{Im}[M(t)]-\mathrm{Im}[M^T(t)])$ and $M(t)=\sum_il_i\otimes l_i^*$. The matrix $\mathcal{H}(t)$ corresponds to single particle Hamiltonian in Majorana basis while $M(t)$ encodes all the bath information  (see Appendix.~\ref{app_num} for detail) and is therefore time independent in our case. Substituting the ansatz $C(t)=Q(t)C(0)Q^T(t)-iP(t)Q^T$  in the above equation, where $P(T)$ and $Q(t)$ are two real matrices,  results in two simpler equations \cite{prosen11}
\begin{subequations}
	\begin{align}
	\dot{Q}(t)=-X(t)Q(t)
	\end{align}
	\begin{align}
	\dot{P}(t)=-X(t)P(t)-Y(t)Q^{-T}(t)
	\end{align}
\end{subequations}
with $Q(0)=\mathcal{I}$ and $P(0)=\mathbb{O}$.
}

{The uniform and time independent coupling of the bath with the system in our case ($\kappa_i=\kappa$) allows us to rewrite $M(t)=M=\kappa\widetilde{M}$ and $Y(t)=Y=\kappa \widetilde{Y}$ where all elements of matrix $\widetilde{M}$ and $ \widetilde{Y}$, are dimensionless. Assuming natural units, we substitute $t=\tilde{t}\tau$, $\widetilde{\mathcal{H}}=\mathcal{H}\tau$, and $\tilde{\kappa}=\kappa\tau$ to arrive at the non-dimensionalised version of the above equations,
\begin{subequations}
 \begin{align}\label{eq_Q}
 	\dot{Q}(\ttil)=-4\left(i\Htil(\ttil)+\ktil Re[\Mtil]\right)Q(\ttil)
 \end{align}
 \begin{align}\label{eq_P}
 	\dot{P}(\ttil)=-4\left(i\Htil+\ktil Re[\Mtil]\right)P(\ttil)-\ktil \Ytil Q^{-T}(\ttil)
 \end{align}
\end{subequations}
Solving the above pair of matrix equations perturbatively to first order in $\ktil$, we obtain} {(see Appendix.~\ref{app_perturb})}
{\begin{widetext}
\begin{align}\label{eq_Corr}
C(\ttil)=V(\ttil)C(0)V^T(\ttil)-\ktil V(\ttil)\big[C(0)\Lambda^T(\ttil)+\Lambda(\ttil)C(0)-\Gamma(\ttil)\big]V^T(\ttil)+\mathcal{O}(\ktil^2),
\end{align} 
\end{widetext}
where,
\begin{subequations}
\begin{align}
V(\ttil)=\mathcal{T}e^{-4i\int_0^{\ttil}\Htil(\ttil')d\ttil'}
\end{align}	
\begin{align}
\Lambda(\ttil)=4\int_{0}^{\ttil}V^T(\ttil')Re[\Mtil]V(\ttil')d\ttil'
\end{align}
\begin{align}
\Gamma(\ttil)=i\int_{0}^{\ttil}V^T(\ttil')\Ytil V(\ttil')d\ttil'.
\end{align}
\end{subequations}
}
{The end of the ramp protocol corresponds to $\ttil=1$ in the rescaled units; the covariance matrix at the end of the ramp is therefore obtained as
\begin{align}
C(1) = V(1)C(0)V^T(1)-\ktil K+\mathcal{O}(\ktil^2),
\end{align}
where $K=V(1)\big[C(0)\Lambda^T(1)+\Lambda(1)C(0)-\Gamma(1)\big]V^T(1)$ is constant matrix with dimensionless elements. Next, we note that the first term in the above equation $V(1)C(0)V^T(1)$ correspond to an unitary time-evolution of the covariance matrix generated by the Hamiltonian $\Htil$. Further, this term captures the full evolution of the covariance matrix in the absence of bath ($\ktil=0$). We, therefore identify this term as the time-evolved covariance matrix $C^U(1)$ in the absence of any dissipative channels. Returning to the original unscaled units, we finally obtain,
\begin{align}
C(\tau)=C^U(\tau)-\kappa\tau K+\mathcal{O}(\kappa^2\tau^2).
\end{align}
}

Let us now consider the Majorana edge correlation $\mathrm{Tr}(\rho(\tau)\theta)=iC_{1,2L}(\tau)$ at the end of the ramp. As we are interested in the adiabatic limit $L\ll\tau$,  we assume that the unitary ramp (in absence of dissipation) induces negligible excitations on its own, i.e., $iC^U_{1,2L}\simeq\bra{\psi_f^0}\theta\ket{\psi_f^0}$.  Using Eq.~\eqref{eq_chi_mix}, the defect $\chi(\tau)$ is obtained as,
\begin{align}
\chi(\tau)=\kappa\tau K_{1,2L}+\mathcal{O}(\kappa^2\tau^2).
\end{align}
Hence, we conclude that in the limit $L\ll\tau\ll\tau_B$, the defect at the end of the ramp scales as $\chi(\tau)\sim\kappa\tau$ in the leading order of perturbation. Further, we note that the above scaling holds true for all pairwise Majorana correlations or elements of the covariance matrix. Consequently, any observable which is a linear function of two point Majorana correlations (for example, the bulk residual energy reported in [\onlinecite{keck17}]) is also expected to satisfy the same scaling behavior obtained above. {Further, we note that the perturbative solution obtained in Eq.~\eqref{eq_Corr} can be used for
 the quasi-local bath discussed in Sec.~\ref{sec_diss_aid_prep}.}

Finally, we note that the perturbation theory we have developed, in essence, extracts the effect of a weak environmental effect (perturbation). The defect generated because of non-adiabatic effects arising from a unitary finite ramp  duration   can not be captured within this framework.

\section{Discussions and Conclusion~}\label{conclude}
 {Choosing two fermionic dissipative environments of considerable difference in character, we quantitatively explore the pros and cons of dissipative annealing in a finite Kitaev chain.  The issue of  the protection of the correlated Majorana fermions against dissipation and environmental interactions has been clearly addressed in this work with reference to the global nature of different environmental couplings.\\
	
Firstly, the chemical potential of a finite 1D Kitaev chain is slowly quenched across a QCP, to drive it from a topologically trivial to a non-trivial phase. In the  thermodynamic limit, even under unitary dynamics, it is well established that annealing across a QCP necessarily induces excitations in the system which in turn annihilates the correlated edge-Majoranas and they delocalise into the bulk of the chain.
	On the other hand, for a finite chain, in the adiabatic regime, the system does not see a gapless point while crossing the QCP;  exlploiting this we have established that the adiabatic preparation of correlated Majorana fermions is possible in harmony with the topologically non-trivial final bulk Hamiltonian.\\
	
	 We then explore the possibility of dissipative preparation of correlated Majorana edge modes through annealing in an open Kitaev chain. {In the first situation discussed in Sec. \ref{sec_diss_aid_prep}, we consider the action of a specifically engineered quasi-local bath. The same bath was also used in  Ref. [\onlinecite{dhiel11}] which analysed the 
	 fate of the Majorana edge correlations, initially present in the system,  in the final steady state of the system: For $\mu=0$, it was established that  the Majorana edge modes reside in a decoherence free sub-space and is unaffected by the dissipative environment.}  In this work, on the contrary, we start from the trivial phase and track the non-equilibrium emergence of mutually correlated edge Majoranas in the steady state under the action of  such a quasi-local dissipation along with a linear ramping of the chemical potential. We establish that the dissipator takes the system into a steady topologically non-trivial dark state and the dynamical correlation between the edge-Majorana assumes a maximum value asymptotically. Particularly, numerical results reveal that a short but finite ramp duration helps the system speedily achieve a high fidelity to the asymptotic topological steady state. We remark that these results are valid even when the  final chemical potential $\mu_f \neq 0$.

In the second situation, we deal with a bath which is locally coupled individually and independently to each fermionic site of the chain. 
Furthermore, the  bath  is  chosen to act as an infinite fermionic reservoir accounting for local particle loss at every individual site of the chain. 
The end sites of the chain have been chosen not to couple with the bath explicitly, to study the indirect effect of dissipation through interaction of the end sites with the bulk chain.
We found both numerical as well as analytical evidences of  an optimal ramp duration $\tau_o$ for which the defect generated in Majorana correlations at the end of the ramp is minimum. The existence of such an optimality is a consequence of the following fact: while the defect generated by the ramp protocol alone is minimised in the adiabatic limit of long ramp duration, the opposite is true for the dissipative dynamics which produces more defects with increasing ramp duration. 

	

In the regime where the dissipative dynamics dominates over the unitary quench, the non-equilibrium residual energy of the system also scales linearly with dissipation strength while exhibiting a Kibble-Zurek (KZ) behaviour in the unitary dominated regime. 
In the present work however, we focus on the defect in edge-Majorana correlation which unlike the residual energy is  a non-extensive quantity and is not observed to follow a KZ scaling even in early time dynamics. Indeed, as mentioned previously, the scaling is exponential with a non-universal coefficient. Nevertheless, the defects induced by the dissipative effects alone display a universal scaling law $\sim\kappa\tau$ in the adiabatic limit of the ramping protocol. The perturbation scheme we propose in Sec. \ref{sec_perturb} justifies this numerical scaling for both the residual energy observed in Ref. [\onlinecite{keck17}] as well as in edge-Majorana correlations reported in our work. Finally, we reiterate that the optimality discussed in this work is also non-universal as there is no universal scaling of the edge-Majorana correlation even for perfectly unitary ramping. 
 \\
	

The Kitaev chain can be experimentally studied in optical lattices with trapped ultra-cold atomic systems \ct{buhler14,kraus12}. The dissipative baths employed in this work can also be engineered with dual interacting optical lattices coupled to a Bose Einstein condensate reservoir \ct{diehl08}. The experimental study of the claims made in the work will further open up the possibility of dynamical preparation of topological Majorana fermionic modes even in contact with an environment. In future, it will be interesting to explore the dissipative preparation of Majorana correlations through a linear quench across the multi-critical point of the Kitaev chain. \cite{mukherjeeepl10}}


\appendix

\section{Numerical scheme for calculating $\chi(t)$}\label{app_num}
In this appendix, we outline the numerical scheme for calculating the defect $\chi(t)$ generated in the edge Majorana correlations during the dissipative evolution of the system. In general, solving Eq.~\eqref{eq_master} tantamount to solving a differential matrix equation with dimensions $2^L\times2^L$. The maximum system size, whose dynamics can be solved numerically is consequently limited. However, note that the Lindblad operators and the system Hamiltonian chosen in our work are linear and  quadratic, respectively, in Majorana operators, i.e., they can be expressed as
\begin{subequations}
\begin{equation}\label{app_majo_L}
L_j=l_j\cdot a=\sum_k l_{j,k} a_k,
\end{equation}
\begin{equation}\label{app_majo_H}
H(t)=a\cdot \mathcal{H}\cdot a=\sum_{i,j}\mathcal{H}_{i,j}a_ia_j.
\end{equation}
\end{subequations}

{By associating a Hilbert space structure $\hat{O}\to\ket{\hat{O}}$ to the space of operators $\mathcal{K}$ with a canonical basis $\ket{P_\mathrm{\alpha}}$,	
\begin{align}
P_\mathrm{\alpha} = a_1^{\alpha_1}a_2^{\alpha_2}...a_{2n}^{\alpha_{2n}}\quad \alpha_j\in\{0,1\}, 
\end{align}
orthonormal with respect to an inner product $\braket{\hat{O}_1|\hat{O}_2}=2^{-n}\mathrm{tr}(\hat{O}_1^\dagger\hat{O}_2)$, it is possible to 
recast Eq.~\eqref{eq_master} in the form,
\begin{align}\label{app_master}
\frac{\partial\ket{\rho}}{\partial t} = \mathcal{L}\ket{\rho} = b.A.b - A_0\mathcal{I},
\end{align}
where, $b_i$ are Majorana operators defined over the space $\mathcal{K}$. The matrices $A$ and $A_0$ are defined as 
\begin{subequations}\label{app_A}
\begin{align}
A_{2j-1,2k-1}&=-2iH_{jk}-M_{jk}+M_{kj}\nonumber\\
A_{2j-1,2k}&=2iM_{kj}\nonumber\\
A_{2j,2k-1}&=-2iM_{jk}\nonumber\\
A_{2j,2k}&=-2iH_{jk}+M_{jk}-M_{kj},
\end{align}
\begin{align}
A_0=2\mathrm{tr}(M),
\end{align}
\end{subequations}
where $M=\sum_il_i\otimes l_i^*$. $A$ is an antisymmetric matrix which implies that its eigenvalues always come in pairs $\beta,-\beta$.} 

{\paragraph{Time-scale of relaxation to steady state ($\tau_B$)}
For a time-independent Hamiltonian, one can show that the steady state corresponds to the zero eigenvalue of $\mathcal{L}$ defined in Eq.~\eqref{app_master}, while the rate of relaxation to steady state is given by $\tau_B = min\{Re[\beta_i]\in\mathcal{Z^+}\}^{-1}$. As an illustration, let us consider the case of a Kitaev chain with PBC, evolving under a constant Hamiltonian with the Lindblad operators chosen as in Eq.~\eqref{eq_lindblad_local}. Eq.~\eqref{eq_master} then decouples into a set of $L/2$ $4\times4$ dimensional independent matrix equations, each corresponding to a quasi-momentum $k$,
\begin{align}
\frac{\partial\rho_k}{\partial t}=-i[H_k,\rho_k(t)]+\kappa\big(2c_k\rho(t)c_k^\da-\{c_k^\da c_k,\rho(t)\}\nonumber\\+2c_{-k}\rho(t)c_{-k}^\da-\{c_{-k}^\da c_{-k},\rho(t)\}\big).
\end{align}   
where, $H_k$ is defined in Eq.~\eqref{eq_hamil_k}. Rewriting $H_k$ and $c_{k,-k}$ as in Eq.~\eqref{app_majo_H} and .~\eqref{app_majo_L}, respectively, one can construct the $8\times8$ dimensional matrix $A$ using Eq.~\eqref{app_A}. The eigenvalues of $A$, as already mentioned, can be divided into pair of two sets differing by their sign. One can check that the set of unique eigenvalues for the $8\times8$ dimensional matrix $A$ is given by \{$\kappa+iE_k$, $\kappa+iE_k$, $\kappa-iE_k$, $\kappa-iE_k$\}, where $E_k$ is defined in Eq.~\eqref{eq_spec}. The time-scale of relaxation to steady state is hence obtained as $\tau_B\sim1/\kappa$}.
{\paragraph{Dynamical equation for Majorana edge correlations}
For the Kitaev chain with OBC, once can construct a} covariance matrix $C_{i,j}(t)=\Tr(a_ia_j\rho(t))-\delta_{i,j}$  of dimension $2L\times 2L$ that encodes all the pair-correlations in the non-equilibrium state of the system \ct{prosen08, prosen10, prosen11}. The { dynamical evolution of} the covariance matrix is obtained numerically by solving the equation \ct{prosen08, prosen10, prosen11}
\begin{align}\label{eq_app_cov}
\dot{C}(t)=-X(t)C(t)-C(t)X^T(t)+iY(t),
\end{align}
where, $X(t)=4(i\mathcal{H}(t)+\mathrm{Re}[M(t)])$, $Y(t)=4(\mathrm{Im}[M(t)]-\mathrm{Im}[M^T(t)])$ and $M(t)=\sum_il_i\otimes l_i^*$. The defect in edge Majorana correlation is then calculated as 
\begin{equation}
\chi(t)=\bra{\psi_f^0}\theta\ket{\psi_f^0}-iC_{1,2L}(t).
\end{equation}

\section{Asymptotic steady state of the system for Lindblad operators chosen as $L_j=d_j^f$}\label{app_steady}
In this appendix, we illustrate that the asymptotic steady state of the system for the Lindblad operator $L_j=d_j^f$ (discussed in Sec.~\ref{sec_diss_aid_prep}) corresponds to the topological ground state of the final Hamiltonian. On substituting $\rho_{ss}=\ket{\psi_f^0}\bra{\psi_f^0}$ in the l.h.s of Eq.~\eqref{eq_steady}, the first term trivially vanishes as $\ket{\psi_f^0}$ is an eigen (ground) state of $H_f$. Further, by construction, the operators $d_j^f$ annihilates the state $\psi_f^0$. 
\begin{equation}
d_j^f\ket{\psi_f^0}=\bra{\psi_f^0}d_j^{f\dagger}=0.
\end{equation}
Consequently, the second term in the l.h.s of Eq.~\eqref{eq_steady} also vanishes. Therefore, $\rho_{ss}=\ket{\psi_f^0}\bra{\psi_f^0}$ is indeed an asymptotic steady state of the system when the Lindblad operators are chosen as $L_j=d_j^f$. In addition, we have verified numerically through an eigenvalue analysis of the Lindbladian super operator \cite{prosen08} that the steady state is also unique.

{\section{Perturbative solution to the equation of motion for the covariance matrix}\label{app_perturb}
Recalling Eq.~\eqref{eq_Q}, we have,
\begin{align}\label{app_Q}
 	\dot{Q}(\ttil)=-4\left(i\Htil(\ttil)+\ktil Re[\Mtil]\right)Q(\ttil).
\end{align}
To facilitate a perturbative expansion, we use the transformation,
\begin{align}
Q_I(\ttil)=iV^T(\ttil)Q(\ttil),
\end{align}
to rewrite Eq.~\eqref{app_Q} as,
\begin{align}\label{app_QI}
\dot{Q}_I(\ttil)=\left(\dot{V}^T(\ttil)-4iV^T(\ttil)\Htil(\ttil)\right)V^{-T}(\ttil)Q_I(\ttil)\nonumber\\-4\ktil V^T(\ttil)Re[\Mtil]V^{-T}(\ttil)Q_I(\ttil).
\end{align}
}
{We now demand that $\dot{V}^T(\ttil)=4iV^T(\ttil)\Htil(\ttil)$ with $V(0)=\mathcal{I}$, which yields the solution
\begin{align}
V(\ttil)=\mathcal{T}e^{-4i\int_0^{\ttil}\Htil(\ttil')d\ttil'},
\end{align}
where we have made use of the fact that $\mathcal{H}(\ttil)$ is anti-symmetric. The anti-symmetric nature of $\mathcal{H}(\ttil)$ can be easily verified by comparing Eq.~\eqref{eq_hamil_ma} and~\eqref{app_majo_H} and using the anti-commutation properties of Majorana operators. Note that the above solution also implies $V^\dagger(\ttil)=V^{-1}(\ttil)$. However, one can check that the matrix $i\Htil(\ttil)$ is real, which implies $V^T(\ttil)=V^{-1}(\ttil)$. As a result, Eq.~\eqref{app_QI} assumes the form,
\begin{align}
\dot{Q}_I(\ttil)=-4\ktil V^T(\ttil)Re[\Mtil]V(\ttil)Q_I(\ttil).
\end{align}
Using the initial condition $Q_I(0)=iV^T(0)Q(0)=i\mathcal{I}$, we obtain the solution for the above equation to first order in $\ktil$:
\begin{align}
Q_I(\ttil)=i\mathcal{I}-4i\ktil \int_0^{\ttil}V^T(\ttil')Re[\Mtil]V(\ttil')d\ttil'+\mathcal{O}(\ktil^2),
\end{align} 
or,
\begin{align}\label{app_Q_final}
Q(\ttil)=V(\ttil)\left(\mathcal{I}-\ktil\Lambda(\ttil)\right)+\mathcal{O}(\ktil^2),
\end{align}
where we have set,
\begin{align}
\Lambda(\ttil)=4\int_0^{\ttil}V^T(\ttil')Re[\Mtil]V(\ttil')d\ttil'. 
\end{align}
Inverting Eq.~\eqref{app_Q_final} and retaining terms up to first order in $\ktil$, we get
\begin{align}\label{app_Qinv_final}
Q^{-1}(\ttil)=\left(\mathcal{I}+\ktil\Lambda(\ttil)\right)V^T(\ttil)+\mathcal{O}(\ktil^2),
\end{align}
}

{Next, we recall Eq.~\eqref{eq_P},
\begin{align}\label{app_P}
 	\dot{P}(\ttil)=-X(\ttil)P(\ttil)-\ktil \Ytil Q^{-T}(\ttil),
\end{align}
where $X(\ttil)=4\left(i\Htil+\ktil Re[\Mtil]\right)$. We note that, 
\begin{align}
X(\ttil)=-\dot{Q}(\ttil)Q^{-1}(\ttil)=Q(\ttil)\dot{Q}^{-1}(\ttil),
\end{align} 
where the first equality follows directly from Eq.~\eqref{app_Q} while the second equality follows trivially from taking the derivative of the equation $Q(\ttil)Q^{-1}(\ttil)=\mathcal{I}$ with respect to $\ttil$. Substituting the above equation in Eq.~\eqref{app_P} and multiplying the same with $Q^{-1}(\ttil)$ yields,
\begin{align}
\frac{d}{d\ttil}\left(Q^{-1}(\ttil)P(\ttil)\right)=-\ktil Q^{-1}(\ttil)\Ytil Q^{-T}(\ttil),
\end{align}
or,
\begin{align}
P(\ttil)=-\ktil Q(\ttil)\int_0^{\ttil}Q^{-1}(\ttil')\Ytil Q^{-T}(\ttil')d\ttil',
\end{align}
where we have used the condition $P(0)=\mathbb{O}$. Substituting Eq.~\eqref{app_Q_final} and~\eqref{app_Qinv_final} in the above equation, we get
\begin{align}\label{app_P_final}
P(\ttil)=i\ktil V(\ttil)\Gamma(\ttil)+\mathcal{O}(\ktil^2),
\end{align}
where we have defined,
\begin{align}
\Gamma(\ttil)=i\int_{0}^{\ttil}V^T(\ttil')\Ytil V(\ttil')dt'
\end{align}
The covariance matrix is now obtained by substituting $Q(\ttil)$ and $P(\ttil)$ from Eqs.~\eqref{app_Q_final} and~\eqref{app_P_final}, respectively, in
\begin{align}
C(\ttil)=Q(\ttil)C(0)Q^T(\ttil)-iP(\ttil)Q^T(\ttil).
\end{align}
Retaining terms up to first order in $\ktil$, we finally arrive at
\begin{widetext}
\begin{align}
C(\ttil)=V(\ttil)C(0)V^T(\ttil)-\ktil V(\ttil)\left[\Lambda(\ttil)C(0)+C(0)\Lambda^T(\ttil)-\Gamma(\ttil)\right]V^T(\ttil)+\mathcal{O}(\ktil^2)
\end{align}
\end{widetext}
}  

\begin{acknowledgements}
We acknowledge Utso Bhattacharya, Somnath Maity, Arijit Kundu and Sourav Biswas for comments and discussions. Souvik Bandyopadhyay and  Sourav Bhattacharjee  acknowledge CSIR, India for financial support.
Souvik Bandyopadhyay also acknowledges PMRF, MHRD, India.
\end{acknowledgements}

\end{document}